\documentclass[%
aip,
reprint,
superscriptaddress,
amsmath,amssymb,
apl,
longbibliography,
]{revtex4-1}

\usepackage{flafter} 

\usepackage{graphicx}
\usepackage{color}
\usepackage{dcolumn}
\usepackage{upgreek}
\usepackage{bm}
\usepackage[normalem]{ulem}
\usepackage{multirow}
\usepackage {hyperref}

\usepackage[utf8]{inputenc}
\usepackage[T1]{fontenc}
\usepackage{mathptmx}

\begin{document}


	\title{Dispersive readout of reconfigurable ambipolar quantum dots in a silicon-on-insulator nanowire}
	
	\author{Jingyu Duan}
	\affiliation{London Centre for Nanotechnology, University College London, London WC1H 0AH, United Kingdom}
    \affiliation{Quantum Motion Technologies, Nexus, Discovery Way, Leeds, LS2 3AA, United Kingdom}

	\author{Janne S. Lehtinen}
	\affiliation
	{VTT Technical Research Centre of Finland Ltd, P.O. Box 1000, 02044 VTT Espoo, Finland}

    \author{Michael~A.~Fogarty}
	\affiliation{London Centre for Nanotechnology, University College London, London WC1H 0AH, United Kingdom}
	\affiliation{Quantum Motion Technologies, Nexus, Discovery Way, Leeds, LS2 3AA, United Kingdom}

	\author{Simon Schaal}
	\affiliation{London Centre for Nanotechnology, University College London, London WC1H 0AH, United Kingdom}
	\affiliation{Quantum Motion Technologies, Nexus, Discovery Way, Leeds, LS2 3AA, United Kingdom}

	\author{Michelle Lam}
    \affiliation{Department of Physics and Astronomy, University College London, London WC1E 6BT, United Kingdom}

	\author{Alberto Ronzani}
	\affiliation
	{VTT Technical Research Centre of Finland Ltd, P.O. Box 1000, 02044 VTT Espoo, Finland}
	
	\author{Andrey Shchepetov}
	\affiliation
	{VTT Technical Research Centre of Finland Ltd, P.O. Box 1000, 02044 VTT Espoo, Finland}

	\author{Panu Koppinen}
	\affiliation
	{VTT Technical Research Centre of Finland Ltd, P.O. Box 1000, 02044 VTT Espoo, Finland}

	\author{Mika Prunnila}
	\affiliation
	{VTT Technical Research Centre of Finland Ltd, P.O. Box 1000, 02044 VTT Espoo, Finland}

    \author{Fernando Gonzalez-Zalba}
    \affiliation{Quantum Motion Technologies, Nexus, Discovery Way, Leeds, LS2 3AA, United Kingdom}

    \author{John J. L. Morton}
    \affiliation{London Centre for Nanotechnology, University College London, London WC1H 0AH, United Kingdom}
    \affiliation{Quantum Motion Technologies, Nexus, Discovery Way, Leeds, LS2 3AA, United Kingdom}
    \affiliation{Department of Electronic \& Electrical Engineering, University College London, London WC1E 7JE, United Kingdom}

\date{\today}
	
\begin{abstract}
	
We report on ambipolar gate-defined quantum dots in silicon on insulator (SOI) nanowires fabricated using a customised complementary metal-oxide-semiconductor (CMOS) process. The ambipolarity was achieved by extending a gate over an intrinsic silicon channel to both highly doped n-type and p-type terminals. We utilise the ability to supply ambipolar carrier reservoirs to the silicon channel to demonstrate an ability to reconfigurably define, with the same electrodes, double quantum dots with either holes or electrons. We use gate-based reflectometry to sense the inter-dot charge transition(IDT) of both electron and hole double quantum dots, achieving a minimum integration time of 160(100) $\upmu$s for electrons (holes). Our results present the opportunity to combine, in a single device, the long coherence times of electron spins with the electrically controllable holes spins in silicon. 
	
\end{abstract}
	
\maketitle
	
The spin degree of freedom of single electrons bound to quantum dots in silicon is considered  one of the most scalable candidates to host quantum information~\cite{loss.ea1998quantum}. By isotopic purification of the material, the Hahn-echo coherence time has been extended up to 28 ms~$^[$\cite{veldhorst.ea2014addressable}$^]$, enabling magnetically-driven single and two-qubit control fidelities of over 99.9$\%^[$\cite{Yang2019silicon}$^]$ and 98$\%^[$\cite{Huang2018fidelity}$^]$, respectively. 
All-electrical control of spin qubits via the spin-orbit interaction can be used to achieve faster and more scalable control, however, the intrinsic spin-orbit coupling of electron spins are too weak to induce high-fidelity coherent rotations~\cite{Corna2017electrically}. In contrast, hole spins are subject to stronger spin-orbit fields, enabling fast two-axis control of the qubit albeit with the drawback of sub-microsecond coherence times~\cite{maurand.ea2016cmos,watzinger.ea2018germanium,zwanenburg2009spin,hendrickx.ea2019fast,liles2020electric}. 

Electron and hole spin qubits have typically been achieved using using different host materials and/or gate stacks. Ambipolar devices, able to operate in both electron and hole regimes, are interesting platforms to attempt to combine the best features of both and to explore their performance within the same crystalline environment~\cite{mueller.ea2015electron-hole,mueller.ea2015single-charge}. 
Ambipolar transport has been previously demonstrated in group IV materials such as graphene~\cite{guttinger2009electron}, carbon nanotubes~\cite{Laird2015,Jarillo-Herrero2004,Pei2012} and germanium~\cite{li2006study}. In silicon MOS devices, ambipolar quantum dots have been achieved by integrating both n-type and p-type reservoirs in a single device~\cite{Prunnila2008,betz.ea2014ambipolar,mueller.ea2015single-charge,mueller.ea2015electron-hole,spruijtenburg2016passivation}, or by tuning the reservoir Fermi energy using NiSi source/drain electrodes~\cite{kuhlmann.ea2018ambipolar}. 
Such ambipolar quantum dots have been studied via direct electrical transport and recently, ambipolar charge sensing via single-electron and single-hole charge sensors has been demonstrated~\cite{sousa-de-almeida.ea2020ambipolar}. However, readout via gate-based sensors~\cite{urdampilleta.ea2018gate-based} or direct dispersive readout via spin projection in double quantum dots~\cite{west.ea2018gate-based,pakkiam.ea2018single-shot,zheng.ea2019rapid} offer more compact and scalable measurement methodologies with comparable measurement sensitivity and shorter integration time.

In this Letter, we present a silicon nanowire (SiNW) multiple quantum dot device, fabricated with a double poly-silicon gate layer technology, together with ambipolar carrier reservoirs for supply of either electrons or holes. We demonstrate reconfigurable single and double quantum dots in both n-type and p-type regimes via gate-based dispersive readout~\cite{ahmed.ea2018radio-frequency}. We also discuss the signal-to-noise ratio (SNR) in the detection of inter-dot electron and hole charge transitions, finding minimum integration times, for SNR = 1, of 160 and 100~$\upmu$s, respectively.

\begin{figure}
\centering
\includegraphics[width=1\linewidth]{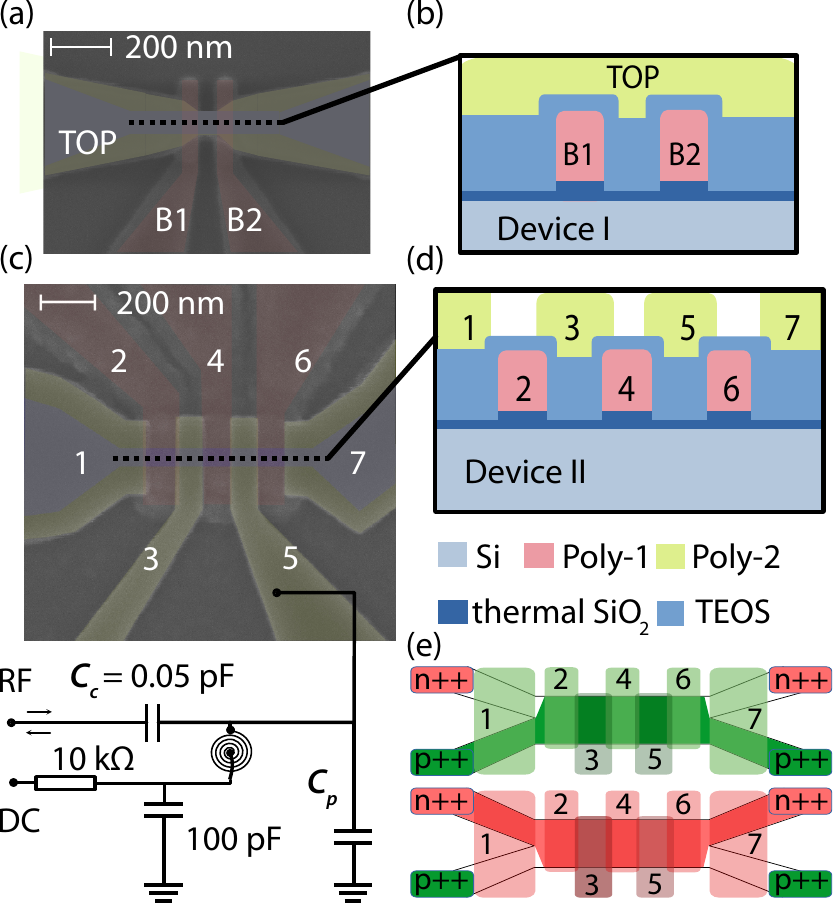}
\caption{(a,c) False colored scanning electron microscope image of devices nominally identical to Device I and Device II. Gate 5 of Device II is attached to an LC circuit for dispersive readout. (b,d) Cartoon cross-sections of the stacked silicon channel, oxide and poly-silicon gates along the dashed line in (a,c); and (e) schematic of ambipolar device operation mode, with accumulation of electron or holes depending on the applied voltage on all the seven gates. Gates 1 and 7 extend from the implanted regions to the channel while gates 2-6 define the quantum dots which confine single electron or holes.} 
\label{fig:Fig1}
\end{figure}

    
Figures~\ref{fig:Fig1}(a-d) show false-colored scanning electron microscopy images and schematic cross-sections of the two types of device studied here, hereinafter named Device I and Device II. The devices were fabricated on 150~mm silicon-on-insulator (SOI) wafers with a customised CMOS process in VTT’s Micronova cleanroom facilities. The process consisted of 8 UV and 3 e-beam lithography layers. The SOI layer was thinned down to 35 nm by thermal oxidation and oxide stripping and patterned to form the nanowires. A 20~nm thermal SiO$_2$ was grown to provide the insulator between the SiNWs and first gate layer. This step reduced the Si layer to its final thickness of 24 nm. The first and second polycrystalline silicon (polysilicon) gate layers, respectively, Poly-1 and Poly-2, have thicknesses 50~nm and 80~nm and were degenerately doped with low energy phosphorous ion implantation. The 35~nm thick SiO$_2$  dielectric layer between the polysilicon gate layers was grown by low-pressure chemical vapour deposition (LPCVD). From van der Pauw structures, we measured the room temperature resistivities of Poly-1 and Poly-2 films to be $\rho_1 = 1.14\times10^{-2} \: \Omega \,$cm and $\rho_2 =1.9\times10^{-3} \: \Omega \,$ cm, respectively. Openings through the deposited dielectrics were etched on the source/drain regions of the SOI and phosphorous (n-type) or boron (p-type) implantation was used to dope these regions. A 250~nm thick SiO$_2$ was deposited with LPCVD and the wafers were heated to 950 $^{\circ}$C to activate the dopants and anneal the dielectrics. Contact holes for all three layers were etched with subsequent dry and wet etching processes. Finally, metallisation layer consisting of 25 nm TiW and 250 nm AlSi was deposited and patterned, and the wafers were treated with a forming gas anneal passivation.

The two devices measured here both have an effective SiNW cross-section of 24~nm $\times$ 24~nm. 
Device I consists of three polysilicon gates: two in Poly-1, with a gate length of 50~nm and pitch of 100~nm, and one Poly-2 that covers the SOI area from source to drain. Device II consists of seven gates for the operation of the ambipolar quantum dots. Extension gates 1 and 7 are used to accumulate carriers in the the intrinsic silicon connecting the quantum dot ``channel'' area to the reservoirs. By applying a positive voltage above some threshold $V_\text{e,th}$, we induce a 2-dimensional electron gas (2DEG) in the channel, supplied by the n-type reservoir contact. Conversely, by applying a negative bias below $V_\text{h,th}$, we induce a 2-dimensional hole gas (2DHG) from the p-type reservoir contact, as illustrated in Fig.~\ref{fig:Fig1}(e). In contrast to the single topgate found in Device I, the distinct gates 1 and 7 present in Device 2 allow for independent control of the left and right reservoir polarity. Gates 2-6 are used to confine quantum dots and tune tunnel coupling between the dots and to the reservoirs. Gates 2, 4, 6 (Poly-1) wrap around the SiNW and have a gate length of 110~nm. Gates 3 and 5 (Poly-2) have a gate length of 120~nm and nominally overlap the Poly-1 gates by 10~nm. 

We use gate-based dispersive readout to sense the charge state of single and double quantum dots in these ambipolar devices. Gate 5 is connected to an LC resonant circuit, consisting of a planar spiral NbTiN superconducting inductor on silicon for high sensitivity reflectometry readout. The choice of gate here is motivated by the much lower resistivity for the Poly-1 versus Poly-2 gates, leading to better high-frequency performance, despite the expected lever arm from this gate on the quantum dots being lower. Together with the parasitic capacitance in the circuit, we obtain a resonance at 489.8~MHz with a resonant bandwidth of 1.64~MHz and a loaded quality factor $Q~=~300$ and a coupling coefficient $\beta=0.33$. The NbTiN thin film thickness is 45~nm and we estimate the total kinetic and geometric inductance of the spiral inductor to be 132~nH~$^[$\cite{mohan.ea1999simple}$^]$. The parasitic capacitance is around 0.8~pF and a surface-mount capacitor of 0.05~pF was used to decouple the resonator from the line~\cite{ahmed.ea2018radio-frequency}. All measurements were conducted at the dilution refrigerator base temperature of 10~mK.

\begin{figure}
\centering
\includegraphics[width=1\linewidth]{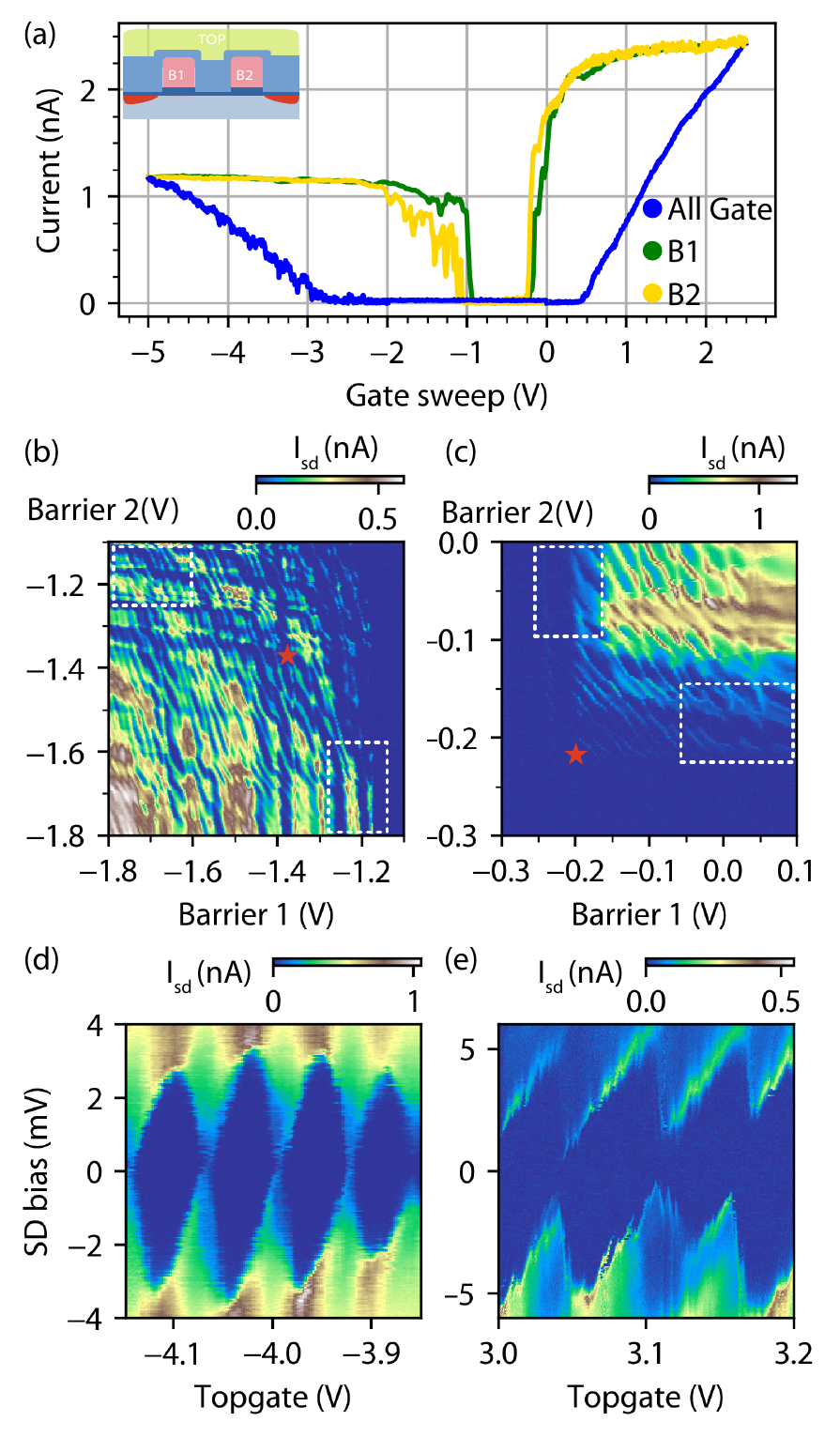}
\caption{Quantum dot formation in Device I: (a) Transport characteristics of Device I, sweep of all gate kept at same voltage(blue), sweep of barrier gate (B1,B2) while other gates kept at well above threshold voltage(2.5V for n-type and -5V for p-type), inset shows the schematic of double barrier gated nanowire(Device I). (b) p-type transport current as a function of each barrier gate B1,B2 at $V_\text{Top} = -4.5V$. (c) n-type transport current as a function of each barrier gate B1,B2 at $V_\text{Top} = 3V$. (d) transport measurement of p-type channel taken at B1 = -1.38V, B2 = -1.38V($\star$ in (b)),(d) transport measurement of n-type channel taken at B1 = -0.20V, B2 = -0.22V($\star$ in (c))} 
\label{fig:Fig2}
\end{figure}

\begin{table}
\begin{ruledtabular}
\caption{\label{tab:table1} Electrostatic properties of the ambipolar quantum dots. }
\begin{tabular}{ccccccc}
\multicolumn{1}{l}{Device} & {QD}  
 & \begin{tabular}[c]{@{}c@{}}$E_c$ \\ (meV)\end{tabular}
 & \begin{tabular}[c]{@{}c@{}}$C_g$ \\ (aF)\end{tabular}
 & \begin{tabular}[c]{@{}c@{}}$C_s$ \\ (aF)\end{tabular}
 & \begin{tabular}[c]{@{}c@{}}$C_d$ \\ (aF)\end{tabular}
 & $\alpha$ \\
\multirow{2}{*}{I (Topgate)}          
& electron 
&   5.4     &   2.2    &   25       &  3         &   0.074\\
& hole     
&   3.2       &    2.4      &   30       &   18       &   0.048    \\
\multirow{2}{*}{II (Gate 5)}       
& electron 
&  17.4        &    1.7      &    4.0      &    3.5      &   0.18    \\
& hole     
&  10.6        &    2.6      &    12      &   0.4     &  0.17     \\
\end{tabular}
\end{ruledtabular}
\end{table}

\begin{figure}
\centering
\includegraphics[width=1\linewidth]{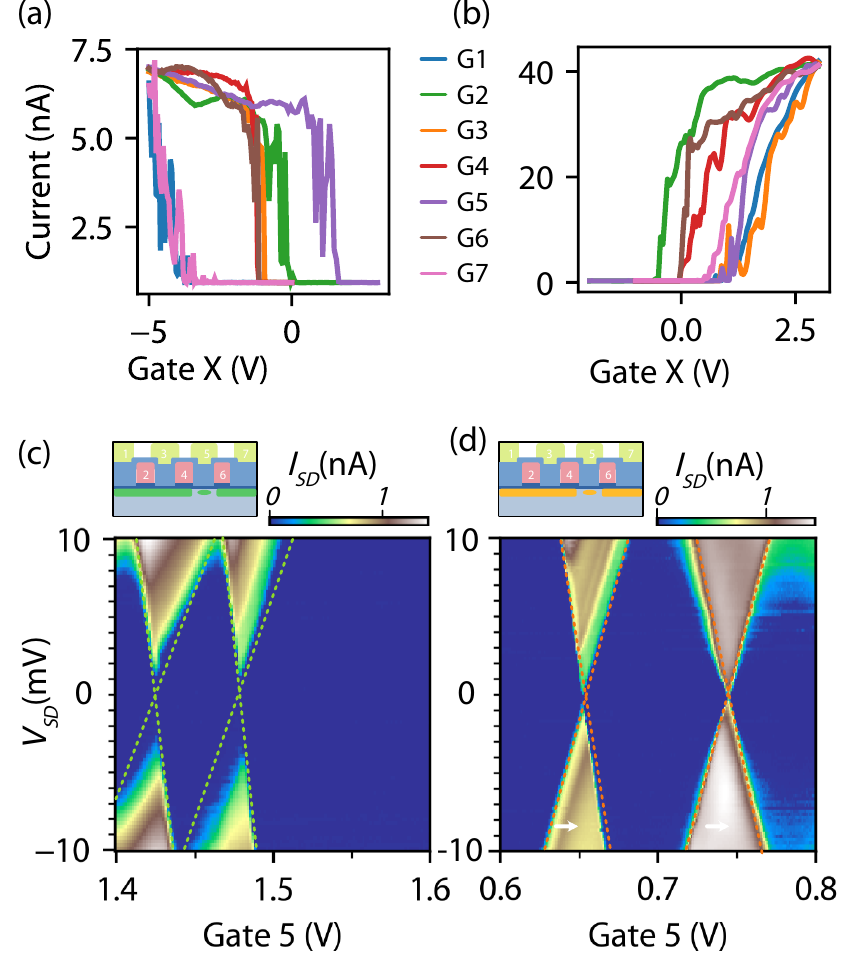}
\caption{ Ambipolar transport Device II: (a) p-type transport measurement ($V_\text{SD} $= 2 mV) as a function of each gate where all other gates were biased at -4.5V. (b) n-type transport measurement ($V_\text{SD} $= 2 mV) as a function of each gate where all other gates were biased at 3V. (c) p-type source to drain current $I_\text{SD}$ as function of Gate 5 and $V_\text{SD}$when other gates were biased at -4.5V. (d) n-type source to drain current $I_\text{SD}$ as function of Gate 5 and $V_\text{SD}$ when other gates were biased at 3V.}
\label{fig:Fig3}
\end{figure}

\begin{figure}
\centering
\includegraphics[width=1\linewidth]{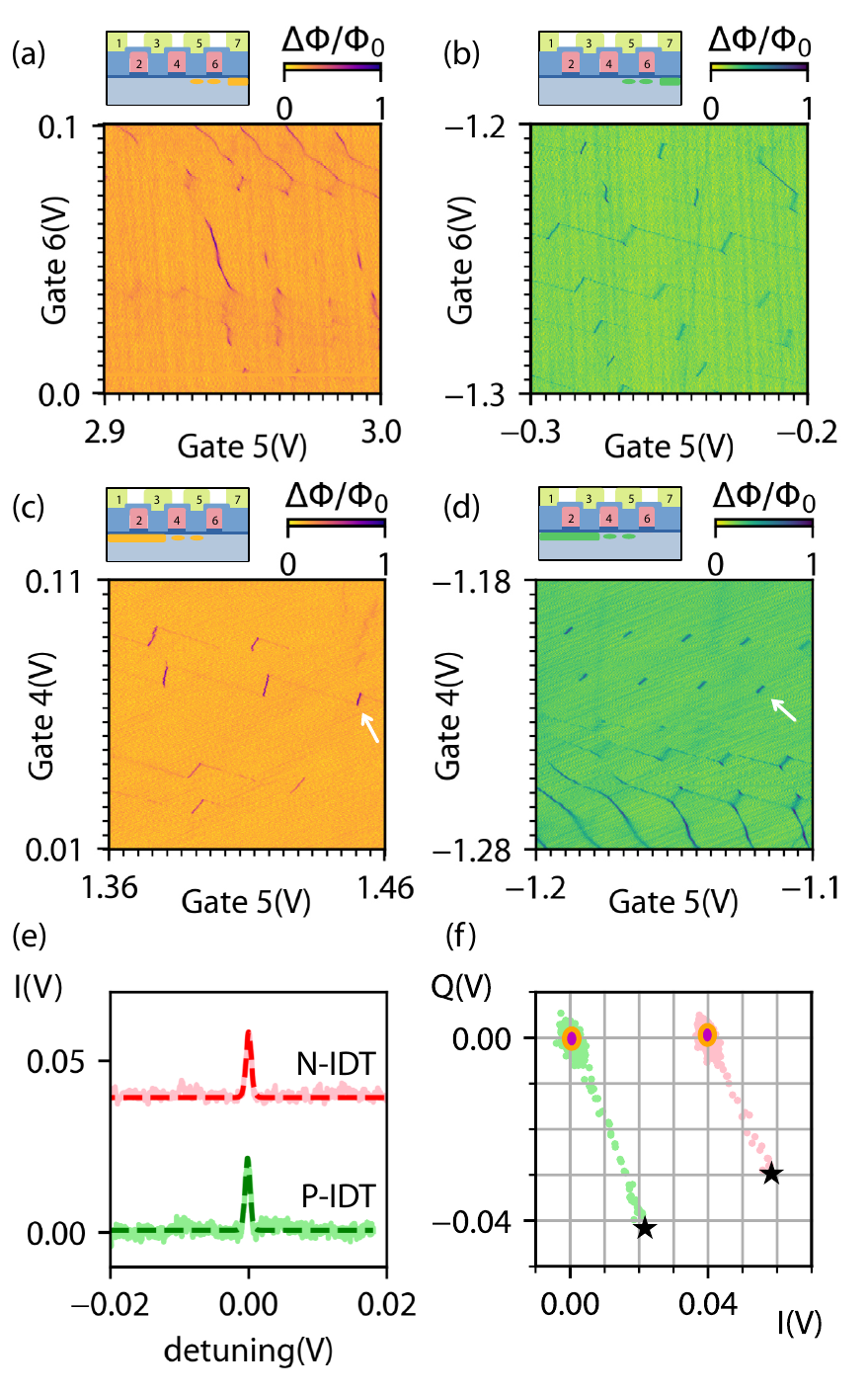}
\caption{Ambipolar double dots: (a-d) different ambipolar double dot configurations and its corresponding charge stability diagrams (a): electron double dots under gate 5 and 6,(b): hole double dots under gate 5 and 6, (c): electron double dots under gate 4 and 5, (d): hole double dots under gate 4 and 5. (e) line trace and fit of homodyne quadrature signal I across IDT in n-type double dot (red, shift up by 0.04V) and p-type(green) measured with input power $P_c$ = -92$\:$dBm. (f) scattered points of quadrature signals in I-Q plane, signal peak ($I_s$,$Q_s$) of the line-fit($\star$), 2D standard deviation of IQ signal.
The demodulated IQ signal are filtered at a low-pass filter of 10kHz and averaged over 300 traces.}
\label{fig:Fig4}
\end{figure}

We first study the quantum dot formation in the SOI channel by measuring the source-drain current of Device I as illustrated in Fig.~\ref{fig:Fig2}(a). Both n-type and p-type transport currents are measured with a source-drain bias voltage $V_\text{SD}$ = 2~mV applied across the source and drain contacts. Topgate threshold voltages for n-type and p-type conduction are measured to be $V_\text{e,th} = 0.40~$V and $V_\text{h,th} = -2.72~$V. 
The barrier gates B1 and B2 have much lower threshold voltages $V_\text{e,th} \simeq -0.2$~V and $V_\text{h,th} \simeq -1~$V over the SOI channel because of the comparatively thinner gate oxide. We investigate effect of the individual barrier gates on the electrical transport in Fig.~\ref{fig:Fig2}(b,c), including their use to form a quantum dot in the silicon channel. Current peaks with a diagonal slope (see red stars) are attributed to a quantum dot formed between two barrier gates, coupled similarly to B1 and B2. Quantum dots can also form under the B1 and B2 gates themselves, thanks to the natural confinement from the silicon nanowire, as can be seen in the vertical and horizontal current peaks (white boxes). From charge stability measurement at fixed barrier voltages, we observe regular Coulomb diamonds corresponding to the central quantum dot in both electron and hole regimes,  with respective charging energies $E_\text{C,e} \simeq 5.4~\text{meV}$ and $E_\text{C,h} \simeq 3.2~\text{meV}$. From these measurements we can extract capacitance values and gate lever arms for electron and holes, as summarised in Table.~\ref{tab:table1}.
The gate capacitance values are consistent with a nominal estimate of 2.5~aF based on a parallel-plate capacitor simplification, with total area $50\times24\times3$~nm$^2$  (considering three sides of the nanowire) and stated oxide parameters --- this suggests that these highly-occupied quantum dots are distributed across most of the SiNW cross-sectional area, as opposed to being localised within the SiNW corners.


Device II was similarly measured in transport and using gate reflectometry. Each gate was confirmed to pinch-off the channel (see Fig.~\ref{fig:Fig3}(a,b)), while the Coloumb diamonds shown in Fig.~\ref{fig:Fig3}(c,d) indicate the formation of electron (hole) quantum dots under gate 5, having been measured with all other gates biased well above (below) the threshold voltage of 3~V ($-4.5$~V).  
The measured lever arms and gate capacitances for this ambipolar quantum dot under gate 5 are presented in Table~\ref{tab:table1}.
These Coulomb diamonds are measured in the few-carrier regime, and correspondingly, the dot-lead capacitance values are much smaller than for the highly-occupied quantum dots studied in Device I. As a result, the gate capacitance dominates and the gate lever arms $\alpha$ are larger. Given the nominal 110~nm gate length in Device II, the measured gate capacitances indicate a smaller effective area of the quantum dot, suggesting that these few-carrier dots are now localised in the top corners of the SiNW cross-section. Similar measurements (shown in Supp.Fig S2) using gate 4 --- which is located in the Poly-1 layer, with much thinner oxide --- yield a larger lever arm of $\alpha_{e} = 0.57$. 


Finally, we demonstrate reconfigurable ambipolar double quantum dots and measure them dispersively. In Fig.~\ref{fig:Fig4}(a--d), we present multiple ambipolar double quantum dots scenarios: Double electron or hole quantum dots located either under gates 5 and 6 (with the source reservoir off), or under  gates 4 and 5 (with the drain reservoir off).

The stability diagram for each scenario is measured by monitoring the normalized phase difference, $\Delta\Phi/\Phi_0$, between the incoming and outgoing radio-frequency signals from the resonator, where $\Phi_0$ is the maximum phase difference. The IDTs within the pair of quantum dots are visible in all four different configurations. No IDTs were observed between dots formed under non-adjacent gates --- the tunnel barriers formed under gates 2, 4 or 6 were evidently too opaque due to their length. 
The magnitude of the dispersive response at the IDT is important for spin readout based on Pauli spin blockade since it determines the maximum signal~\cite{west.ea2018gate-based,zheng.ea2019rapid,pakkiam.ea2018single-shot}. In Fig.~\ref{fig:Fig4}(e), we take two line traces of the IDT reflectometry signal from both electron double quantum dots and hole double quantum dots, illustrated by the arrows in Fig.~\ref{fig:Fig4}(c,d), and filter the demodulated quadrature and in-phase signals with a notch-filter at 16~kHz to suppress a noise peak attributed to the audio component of the pulse tube of the dilution refrigerator~\cite{Kalra2016}. 
By plotting these detuning-dependent traces in ($I$(V), $Q$(V)) space, as shown in see Fig.~\ref{fig:Fig4}(f), the dispersive peak at ($I_s$,$Q_s$) can be identified in the complex plane, facilitating the extraction of the SNR ~\cite{House2015radio,ahmed.ea2018primary}.
We calculate the SNR as $ (I_s-I_0)^2+(Q_s-Q_0)^2\over2\sigma_S^2 $, where $I_0$ and $Q_0$ are respectively the in-phase and quadrature components of the signal far from the IDT and $\sigma_\text{s}$ is the average 2D standard deviation of the noise which can be seen as the radius of the dot around the noise background in the inset of Fig.~\ref{fig:Fig4}(e). We obtain SNR$_\text{e,IDT}$~=~49.8~$\:$dB and SNR$_\text{h,IDT}$~=~52.9~$\:$dB, indicating that our measurement configuration should provide a minimum integration time, for SNR~=~1, of $\tau_\text{e}~=~160~\upmu$s and $\tau_\text{h}~=~100~\upmu$s for electron and holes respectively. 
This sensitivity could be further enhanced by performing reflectrometry using a gate in the Poly-1 layer: the larger lever arms of such gates should give an improvement factor of  $\left(\frac{\alpha_\text{e,Poly-1}}{\alpha_\text{e,Poly-2}}\right)^2\simeq9$.
Operating at a higher reflectometry frequency (e.g.\ 1.8~GHz) should yield a $\sim5\times$ SNR improvement due to reduced parasitic capacitance~\cite{Ibberson2020large}, while further improvements using a Josephson parametric amplifier~\cite{Schaal2020fast} and critically coupled resonator could reduce the integration time down to O(100)~ns. 

In conclusion, we have fabricated and experimentally demonstrated reconfigurable ambipolar quantum dots in a SOI multi-gate nanowire transistor. This work demonstrates the core ingredients necessary to benchmark electron and hole spin qubits in the same silicon device, including RF readout of the IDT which is the basis of Pauli-blockade based spin measurement. Furthermore, the availability of gate-based reflectrometry opens up the possibility of new types of studies in such ambipolar silicon devices. In silicon, ambipolar p-n double quantum dot formation is challenging to measure through direct transport current due to the large silicon bandgap ---  RF readout of the dot-lead charge transition can enable neighbouring quantum dots to be sensed as in the capacitive shift of sensor transition, at zero source-drain bias~\cite{Duan2020Remote}. Furthermore, the same dot-lead charge transition detected by RF reflectometry can provide an accurate measure of the reservoir temperature~\cite{ahmed.ea2018primary}, enabling comparative studies of the electron and hole interactions with phonons within the same nanostructure. 

\begin{acknowledgments}
We thank Gavin Dold and Oscar Kennedy for support and helpful discussions in the fabrication of superconducting inductors. The authors gratefully acknowledge the financial support from the European Union's Horizon 2020 research and innovation programme under grant agreements No 688539 (http://mos-quito.eu) and No 766853 (http://www.efined-h2020.eu/); as well as the Engineering and Physical Sciences Research Council (EPSRC) through the Centre for Doctoral Training in Delivering Quantum Technologies (EP/L015242/1), QUES$^2$T (EP/N015118/1), the Hub in Quantum Computing and Simulation (EP/T001062/1) and Academy of Finland project QuMOS (project numbers 288907 and 287768) and Center of Excellence program project No 312294.
\end{acknowledgments}

%


\end{document}